# Testing General Relativity on the Largest Scales in the Years 1915-1955: the Dawning of Modern Cosmology

Cormac O'Raifeartaigh

[a]*School of Science and Computing, Waterford Institute of Technology, Cork Road, Waterford, Ireland*
Author for correspondence: coraifeartaigh@wit.ie

**Abstract**

Soon after he proposed three classic observational tests for the general theory of relativity, Einstein realised that a consistent description of the universe as a whole provided another important challenge for the theory. In this essay, we recall how general relativity found important application in the nascent field of cosmology, from the static models of Einstein and de Sitter to the expanding cosmologies of Friedmann and Lemaître. By the early 1930s, the first tentative astronomical evidence of cosmic expansion had trigged a paradigm shift to dynamic cosmologies, almost all of which were couched firmly within the framework of general relativity. However, further progress was impeded for some years by a paucity of observational data. In the 1950s, the debate between steady-state and big bang models of the universe provided a new stimulus for the quest to determine key parameters of relativistic models of the cosmos by astronomical observation, setting the stage for today's precision cosmology.

# 1. Introduction

In his first papers outlining the basic principles of the general theory of relativity, Einstein proposed three astronomical tests for the theory – the perihelion of Mercury, the deflection of distant starlight by the sun, and gravitational redshift (Einstein 1915a, 1916). By 1920, many physicists accepted that the first two of these 'classical tests' offered empirical support for the new theory (Eisenstaedt 2006 pp 189-190), while the first tentative observations of gravitational redshift were reported in 1925 (Hentschel, this volume). However, all three tests concerned minor deviations from the predictions of Newtonian gravity, leaving room for some doubt due to uncertainties in observation. Indeed, it is generally accepted that general relativity entered a 'low watermark' period of hibernation after these early years, mainly because of difficulties in connecting with experiment (Eisenstaedt 1989).

Cosmology, the study of the universe on the largest scales, constituted an important exception during this period. As Einstein himself quickly realised, a consistent description of the universe as a whole posed an important challenge for the general theory of relativity. After all, it was a fundamental tenet of general relativity that the geometric structure of a region of spacetime is determined by the mass and energy it contains; thus considerations of the universe at large formed a natural testbed for the theory. Indeed, it is clear from Einstein's correspondence that cosmic considerations were a major preoccupation for him in the immediate aftermath of the publication of the general theory of relativity (Realdi and Peruzzi 2009; O'Raifeartaigh et al. 2017). As he remarked to the Dutch astronomer Willem de Sitter: *"For me, though, it was a burning question whether the relativity concept can be followed through to the finish, or whether it leads to contradictions. I am satisfied now that I was able to think the idea through to completion without encountering contradictions"* (Einstein 1917a).

Thus, the first relativistic model of the cosmos, known as the Einstein World, appeared within two years of the publication of the general theory of relativity. This paper, along with an alternative cosmic model proposed a few months afterwards by Willem de Sitter, set the stage for cosmology throughout the 1920s, although the latter model constituted a source of much confusion for some years. By the early 1930s, astronomical observations had offered the first tentative evidence of an expansion on cosmic scales, a phenomenon that was greatly mysterious in the context of classical mechanics but readily comprehensible in the context of general relativity. From that point onwards, the stage was set for a revolution in modern cosmology, a revolution that was conducted firmly within the framework of the general theory of relativity.

**2. The static cosmologies of Einstein and de Sitter.**

Einstein's manuscript *'Cosmological Considerations in the General Theory of Relativity'* (Einstein 1917b) was published in February 1917, only two years after his first presentation of the covariant field equations of general relativity (Einstein 1915b). Having first noted some well-known difficulties associated with Newtonian descriptions of the cosmos, Einstein's starting point for the first relativistic model of the universe was the assumption of a static spacetime metric due to a static distribution of matter. As he remarked in the paper: *"The most important fact that we draw from experience as to the distribution of matter is that the relative velocities of the stars are very small as compared with the velocity of light. So I think that for the present we may base our reasoning upon the following approximate assumption. There is a system of reference relatively to which matter may be looked upon as being permanently at rest"* (Einstein 1917b). Most historians agree that this assumption was reasonable at the time; many years were to elapse before the observation of a linear relation between the recession of the distant galaxies and their distance, the first evidence for a non-static universe (Nussbaumer and Bieri 2009 pp 72-74; O'Raifeartaigh et al. 2017 ).

A second assumption was that of a uniform distribution of matter. This assumption implied a universe that was both isotropic and homogeneous on the largest scales and was later named the 'Cosmological Principle'. Assuming also a closed spatial geometry for the cosmos in order to render his model consistent with his views on Mach's principle and the relativity of inertia (Realdi and Peruzzi 2009; O'Raifeartaigh et al. 2017), Einstein found it necessary to add a new term to the general field equations in order to avoid a null solution. This term, which became known as the cosmological constant term, was certainly allowed by relativity; the only constraint was that it had to be small enough that the field equations remained compatible with observations of the motion of the planets in our solar system.

Einstein then showed that, for the case of the universe as a whole, the modified field equations have the solution

$$\lambda = \frac{\kappa\rho}{2} = \frac{1}{R^2} \qquad (1)$$

where $\lambda$ represents the cosmological constant, $\rho$ is the mean density of matter and $R$ is the radius of the cosmos (Einstein 1917b). Thus, the first relativistic model of the cosmos predicted an intuitively satisfactory relation between the size of the universe and the matter it contained. As we have noted elsewhere, it is somewhat surprising that Einstein did not then use equation (1) to extract a rough estimate of the size of the cosmos using astronomical estimates of the density of matter in the Milky Way; his correspondence at the time indicates that he considered such calculations unreliable due to large uncertainties in the latter quantity (O'Raifeartaigh et al. 2017). Instead, in a summary of his paper, Einstein emphasizes that the goal of his paper was to determine whether relativity could give a self-consistent model of the universe:

> Thus the theoretical view of the actual universe, if it is in correspondence with our reasoning, is the following. The curvature of space is variable in time and space, according to the distribution of matter, but we may roughly approximate to it by means of a spherical space. At any rate, this view is logically consistent, and from the standpoint of the general theory of relativity lies nearest at hand; whether, from the standpoint of present astronomical knowledge, it is tenable, will not here be discussed.

*2.1 The de Sitter universe*

In July 1917, the Dutch astronomer and theorist Willem de Sitter noted that the introduction of the cosmological constant term to the field equations of general relativity allowed an alternate cosmic solution, namely the case of a universe with no matter content (de Sitter 1917). Approximating the known universe as an empty universe, de Sitter showed that the field equations could also be satisfied by the solution

$$\rho = 0; \lambda = \frac{3}{R^2} \quad (2)$$

a result he dubbed 'Solution B' to Einstein's 'Solution A' (de Sitter 1917). In this cosmology, Einstein's matter-filled three-dimensional universe of closed spatial geometry was replaced by an empty four-dimensional universe of closed *spacetime* geometry.

Einstein was greatly perturbed by de Sitter's solution, not least because the notion of an empty cosmos was in direct conflict with his understanding of Mach's Principle in these years (Smeenk 2014; O'Raifeartaigh et al. 2017). As he remarked in a paper of 1918: *"It appears to me that one can raise a grave argument against the admissibility of this solution…..In my opinion, the general theory of relativity is a satisfying system only if it shows that the physical qualities of space are completely determined by matter alone. Therefore no $g_{\mu\nu}$-field must exist (that is no space-time continuum is possible) without matter that generates it"* (Einstein 1918a). In the same paper, Einstein suggested a technical objection to de Sitter's model, namely that it contained a spacetime singularity.

In the years that followed, Einstein and de Sitter debated the relative merits of 'Solution A' and 'Solution B' with other theorists such as Cornelius Lanczos, Hermann Weyl, Felix Klein and Gustav Mie. De Sitter's model attracted some interest because of a prediction that light emitted by an object placed in the de Sitter universe would be redshifted, a prediction (known as the 'de Sitter effect') that chimed with emerging observations of the spiral nebulae by the American astronomer V.M. Slipher (Nussbaumer 2013; O'Raifeartaigh 2013). However, it is clear that Einstein never really accepted the de Sitter solution as a realistic model of the universe throughout this debate (Schulmann et al. 1998 pp 351-352). Indeed, it is interesting that Einstein did not refer to de Sitter's model in any of his reviews of cosmology around this time (Einstein 1918b pp 116-118; Einstein 1921; Einstein 1922 pp 110-111).

## 3. Relativistic cosmology in the 1920s

The debate between Einstein, de Sitter and a few other theorists such as Cornelius Lanczos and Hermann Weyl concerning a suitable relativistic model of the cosmos continued throughout the 1920s (Nussbaumer and Bieri 2009 pp 78-83; O'Raifeartaigh et al. 2017). It is probably fair to say that these discussions had almost no impact on the astronomical community, as few astronomers saw any connection between their observations and Einstein's abstruse new theory of gravitation in these years (Kragh 2014). Indeed, if cosmology existed as a field of study at all at this point, it was as a field that was entirely theoretical. Even there, progress was limited because the mathematics of the de Sitter model were not very well understood at the time. In addition, efforts by some theorists to link the de Sitter effect with the well-known redshifts of the nebulae were severely hampered by the fact that the distances of the spirals remained unknown at this point.

An important step forward in theoretical cosmology occurred in 1922, when the Russian physicist Alexander Friedmann suggested that non-static solutions of the Einstein field equations should be considered in relativistic models of the cosmos (Friedmann 1922). Starting from the modified field equations and assuming a positive spatial curvature for the cosmos, Friedmann showed that the theory gave rise naturally to two differential equations linking the time evolution of the cosmic radius $R$ with the mean density of matter $\rho$ and the cosmological constant $\lambda$. Demonstrating that the Einstein and de Sitter models were special cases of this general class of solutions, Friedmann noted that the magnitude of the cosmological constant λ determined whether a matter-filled universe expanded monotonically or expanded and then contracted.

Few physicists paid attention to Friedmann's time-varying cosmology at first, mainly because the work appeared to be quite hypothetical and made no connection with astronomy. Worse, Einstein mistakenly claimed the paper contained a mathematical error and it took over a year for this criticism to be retracted. Even here, an unpublished draft of Einstein's retraction demonstrates that he did not consider Friedmann's cosmology to be realistic.[1]

A few years later, the Belgian physicist Georges Lemaître independently derived time-varying equations for the radius of the cosmos from Einstein's field equations. However, unlike Friedmann, Lemaître made a key connection with astronomy: aware of the well-known redshifts of the spiral nebulae observed by V.M. Slipher and of emerging observations of the extra-galactic distances of the nebulae by Edwin Hubble, Lemaître suggested that the light from these remote astronomical objects was redshifted due to an expansion of space predicted by relativity (Lemaître 1927: Nussbaumer and Bieri 2009 pp 108-110). This work also received very little attention at first, probably because it was published in a little-known Belgian journal and possibly because the redshift/distance relation of the nebulae was not yet well established (O'Raifeartaigh 2020). It is interesting to note that, when Lemaître brought the work to Einstein's attention later that year, the famous physicist dismissed expanding cosmologies as *"abominable"*, a reaction Lemaître attributed to a lack of knowledge of emerging developments in astronomy (Lemaître 1958).

## 3. The paradigm shift to an expanding universe

---

[1] A detailed account of this episode can be found in (Nussbaumer and Bieri 2009 pp 91-92).

In 1929, the American astronomer Edwin Hubble published the first results of a detailed astronomical study of the relation between the redshifts of the nebulae and their radial distance, using redshift data from V.M. Slipher and estimates of nebular distances based on his own observations of individual stars within the nebulae using the 100-inch Hooker telescope at the Mount Wilson Observatory in California (Hubble 1929). The results are reproduced in figure 1. As stated by Hubble: *"the results establish an approximately linear relation between the velocities and distances among nebulae for which velocities have been previously published"*. We note that the distances of at least seven of the nebulae were estimated by observing Cepheid stars within the nebulae and employing Henrietta Leavitt's period-luminosity relation to estimate their distance; this was a key advance on previous methods of estimating the huge astronomical distances of the nebulae. Some commentators have noted that the quality and quantity of the data shown on Hubble's graph only marginally support the conclusion of a linear relation between the velocities (measured as redshifts) and the distances of the nebulae. However, the graph marked an important turning point as the data were accepted as the first evidence of a relation that had long been suspected by astronomers (Smith 1982 pp 180-184). This conclusion was strengthened with the publication of a paper soon afterwards that extended the relation to much larger distances and redshifts (Hubble and Humason 1931; Smith 1982 pp 190-193).

On January 10$^{th}$ 1930, a landmark meeting took place at the Royal Astronomical Society in Burlington House in London. Prominent astronomers such as Arthur Stanley Eddington and Willem de Sitter accepted that the recent observations of a linear relation between the redshifts of the spiral nebulae and their radial distances could not be readily explained in the context of either classic mechanics or of the relativistic cosmic models of Einstein and de Sitter. During the meeting, it was suggested that non-static cosmologies should be considered. This discussion was reported in the February issue of *The Observatory* and read by Georges Lemaître, who immediately wrote a letter to Eddington, reminding him of his own 1927 article on cosmic expansion. As has been detailed elsewhere (Nussbaumer and Bieri 2009 pp 121-126; O'Raifeartaigh 2020), Eddington responded with enthusiasm, publicly acknowledging the importance of Lemaître's 1927 article and bringing it to the attention of his colleagues. Eddington also ensured the article reached a wider audience by arranging for it to be republished in English in the *Monthly Notices of the Royal Astronomical Society* (Lemaître 1931a). Another important development in this period was Eddington's demonstration that Einstein's static model was in any case unstable (Eddington 1930).

Thus by 1931, with the publication of further redshift/distance observations by Hubble and the republication of Lemaître's 1927 paper in English, it seemed to many physicists that the famous redshifts of the nebulae could be satisfactorily explained in terms of a cosmic expansion predicted by the general theory of relativity. This led to the publication of a plethora of dynamic relativistic models of the expanding universe of the Friedmann-Lemaître type with varying values for cosmic parameters such as the curvature of space and the cosmological constant (Nussbaumer and Bieri 2009 pp 137-143). Even Einstein overcame his earlier distrust of time-varying models of the cosmos, declaring during a visit to the Mount Wilson Observatory (figure 2) that: *"New observations by Hubble and Humason concerning the redshift of light in distant nebulae make the presumptions near that the general structure of the universe is not static"* (AP 1931a) and *"The redshifts of the distant nebulae have smashed my old construction like a hammer blow"* (AP 1931b). This period also saw the first articulation of models that envisioned a universe that originated in a compact dense state, expanding outwards ever since (Lemaître 1931b,c).

In April 1931, Einstein published a model of the expanding cosmos based on Friedmann's 1922 analysis of a matter-filled dynamic universe of positive spatial curvature (Einstein 1931). The most important feature of this model, known as the Friedmann-Einstein model, was that Einstein dispensed with the cosmological constant term, for two stated reasons. First, the term was unsatisfactory because it had been shown that it did not in fact provide a stable static solution: "*It can also be shown… that this solution is not stable. On these grounds alone, I am no longer inclined to ascribe a physical meaning to my former solution*" (Einstein 1931). Second, the term was unnecessary because the assumption of stasis was not justified by observation: *"Now that it has become clear from Hubbel's* [sic] *results that the extra-galactic nebulae are uniformly distributed throughout space and are in dilatory motion (at least if their systematic redshifts are to be interpreted as Doppler effects), assumption (2) concerning the static nature of space has no longer any justification*" (Einstein 1931). Most importantly, setting the cosmological constant term to zero allowed Einstein to derive simple expressions relating the rate of cosmic expansion to key parameters such as the present radius of the cosmos, the mean density of matter and the timespan of expansion (Nussbaumer 2014). Using Hubble's empirical estimate of 500 km s$^{-1}$Mpc$^{-1}$ for the recession rate of the nebulae, he calculated numerical values of $10^8$ light-years, $10^{-26}$ g/cm$^3$ and $10^{10}$ years for each of these parameters respectively (Einstein 1931). We have previously noted that these calculations contain a slight systematic numerical error (O'Raifeartaigh and McCann 2014). However, the error did not substantially affect a major puzzle raised by the model; if the timespan of cosmic expansion represented the age of the universe, it was strangely small in comparison with estimates of the age of stars (as calculated from astrophysics) or estimates of the age of the earth (as deduced from radioactivity). Einstein attributed this age paradox to the idealized assumptions of the model, in particular the assumption of a homogeneous distribution of matter on the largest scales (Einstein 1931).

In 1932, Einstein collaborated with Willem de Sitter to propose an even simpler model of the expanding universe that could be compared to astronomical observation. Following an observation by Otto Heckmann that the presence of matter in a non-static universe did not necessarily imply a positive curvature of space, and mindful of a lack of empirical evidence for spatial curvature, Einstein and de Sitter set both the cosmological constant and spatial curvature to zero (Einstein and de Sitter 1932). An intriguing facet of this model was that the rate of expansion and the density of matter were related by the simple expression

$$\left(\frac{R'}{R}\right)^2 = \frac{1}{3}\kappa\rho c^2$$

where $R'/R$ is the fractional rate of expansion, $\rho$ is the density of matter and $\kappa$ is the Einstein constant. Applying once again Hubble's empirical value of $H_0 = 500$ km s$^{-1}$ Mpc$^{-1}$ for the recession rate of the galaxies, the authors found that it predicted a value of 4x10$^{-28}$ g cm$^{-3}$ for the mean density of matter in the cosmos, a prediction they found reasonably compatible with contemporaneous estimates from astronomy.

The Einstein-de Sitter model became very well-known and went on to play a significant role in 20$^{th}$ century cosmology. One reason was that it marked an important hypothetical case in which the expansion of the universe was precisely balanced by a critical density of matter. This allowed for a useful classification of cosmic models; assuming a vanishing cosmological constant, a cosmos of mass density higher than the critical value would be of spherical geometry and eventually collapse, while a cosmos of mass density less than the critical value would be of hyperbolic spatial geometry and expand at an ever-increasing rate. The model also marked an important benchmark case for observers; in the absence of empirical evidence for spatial curvature or a cosmological constant, it seemed the cosmos could be described in terms of just two parameters, the rate of expansion and the density of matter, each of which could be determined independently by astronomy. Indeed, the theory became the standard cosmic model for astronomers for many years, although it suffered from a similar timespan problem as the Friedmann-Einstein model (O'Raifeartaigh, McCann and Mitton 2021).

We note finally that a very different cosmic model of interest was also published in this period. In 1933, the British astrophysicist Edward Arthur Milne published a theory in which the expanding universe was described in the context of Newtonian mechanics, supplemented with the principles of the special theory of relativity (Milne 1933). This model was soon dubbed 'kinematic cosmology' as it was rooted in classical mechanics, took account of the principles of special relativity, but dispensed with the principles of the general theory of relativity. As Milne proposed in his paper: *'The phenomenon of the expansion of the universe is shown to have nothing to do with gravitation, and to be explicable, qualitatively and quantitatively, in terms of a flat, infinite static Euclidean space'* (Milne 1933). In particular, Milne rejected general relativity's prediction that mass and energy could curve spacetime and suggested that the redshifts of the nebulae were recession velocities rather than an expansion of space. Milne's kinematic cosmology became a subject of some debate in the UK for some years, receiving support from physicists such as William McCrea and George McVittie, but most physicists found the theory rather contrived (Nussbaumer and Bieri 2009 158-160). As the years progressed, some mathematicians enjoyed the challenge of perusing quasi-Newtonian models of the expanding universe, but few in the physics community considered such theories realistic models of the universe (Kragh 1996 pp 66-67).

## 4. The period 1940-1955

In contrast to the flurry of activity in the early 1930s, the field of cosmology looked set to stagnate during much of the 1940s due to difficulties in determining accurate estimates of cosmic parameters by observation. To those scientists interested in the field, there seemed little prospect in discriminating between the large variety of cosmological models proposed the previous decade on the basis of empirical observation (Kragh 2014). Indeed, it could be said that cosmology did not truly constitute a *test* for general theory of relativity in these years (Eisenstaedt 1989). Worse, the one cosmic parameter that could be determined with reasonable accuracy – the rate of expansion – continued to imply an age for the universe that was deeply problematic. As Einstein remarked in a key review of cosmology (Einstein 1945 p135):

> Last and not least; The age of the universe in the sense used here, must certainly exceed that of the firm crust of the earth as found from the radioactive minerals. Since determination of age by these minerals is reliable in every respect, the cosmological theory here presented would be disproved if it were found contradict any such results. In this case I see no reasonable solution.

However, two advances occurred in theoretical cosmology during the 1940s that were to have a major bearing on modern cosmology.

*4.1 The hypothesis of primordial nucleosynthesis*

In the late 1940s, the Russian émigré physicist George Gamow suggested that Lemaître's notion of a cosmos with a compact, dense beginning might offer a solution to the puzzle of nucleosynthesis. With the failure of standard models of stellar nucleosynthesis to explain the relative abundance of the lightest chemical elements, Gamow and his colleagues Ralph Alpher and Robert Herman explored whether the phenomenon could be described in the context of nuclear processes in a young universe that was once extremely dense and hot (Gamow 1942, 1946; Alpher, Bethe and Gamow 1948; Alpher and Herman 1948). While this work did not receive much attention at first at first, it later set important constraints on estimates of the mean density of matter in the cosmos. More dramatically, the research program led to the hypothesis that today's cosmos might exhibit a faint background radiation of microwave frequency left over from early epochs (Alpher and Hermann 1948). This strange prediction did not receive much attention at the time, but the serendipitous discovery of the cosmic microwave background in the 1960s was to herald a major milestone in observational cosmology.

*4.2 Steady-state cosmology*

In parallel with the work of the Gamow group, a new type of cosmic model was proposed in the United Kingdom known as the 'steady-state' universe. In this cosmology, the universe expands but remains essentially unchanged in every other respect. Today, the steady-state universe is mainly associated with the theorists Fred Hoyle, Hermann Bondi and Thomas Gold, but other theorists entertained similar ideas (Dubois 2019).

Working together at the University of Cambridge in the late 1940s, Hoyle, Bondi and Gold became sceptical of Lemaître's idea of a fireworks origin for the universe and noted that the simplest evolving cosmologies continued to predict a timespan of expansion that was greatly puzzling. In consequence, the trio explored the idea of an expanding universe that remains essentially unchanged due to a continuous creation of matter from the vacuum. For Bondi and Gold, the idea followed from their belief in the 'perfect cosmological principle', a philosophical principle that proposed that the universe should appear essentially the same to observers in all places *at all times* (Bondi and Gold 1948). More quantitatively, Fred Hoyle constructed a steady-state model of the cosmos using Einstein's field equations with the addition of a 'creation-field' term representing the continuous creation of matter from the vacuum (Hoyle 1948). In many ways, the new term acted like a positive cosmological constant term, giving an exponential expansion of space, almost identical to that of the de Sitter model. A few years later, the British physicist William McCrea proposed a slightly more sophisticated formulation of Hoyle's model, in which the 'creation-field' was replaced by a scalar field on the right-hand side of the field equations, representing a negative pressure (McCrea 1951; Kragh 1996 pp 205-206).

*4.3 The debate beween steady-state and big bang cosmoloies*
As is well known, a significant debate developed during the 1950s and 1960s between the steady-state and 'big bang' models of the cosmos.[2] This debate has been described elsewhere (Kragh 1996); we note here that the debate provided a new stimulus for the attempt to determine key cosmic parameters by astronomical observation (Kragh 2014).

---

[2] It was Hoyle who coined the term for the latter models (Hoyle 1952).

One important development in this period was the opening of the new 200-inch Hale telescope at the Palomar Observatory in California in 1949. During the 1950s, new observations of the spiral nebulae by the American astronomers Walter Baade and Allan Sandage led to successive recalibrations of the great distances of the galaxies; these corrections suggested a smaller Hubble constant, implying a longer timespan of cosmic expansion (Longair 2006 pp 340-342). By the end of the decade, Hubble's original estimate of $H_0$= 500 km s$^{-1}$ Mpc$^{-1}$ had been reduced to about 75 km s$^{-1}$ Mpc$^{-1}$, effectively removing the timespan problem that had dogged relativistic cosmology for so long.

On the other hand, astronomical estimates of the mean density of matter in the universe were beginning to suggest a value far below the critical value of the Einstein-de Sitter model. One method was to count the number of galaxies in a given volume of space and multiply by the mass of each galaxy, the latter figure being obtained by measuring the average luminosity of galaxies and converting to mass by means of the mass-to-light ratio of galactic matter (Longair 2006 pp 357-360). Another method was to determine the mass of galaxies by dynamical methods, i.e., by comparing the motion of rotating galaxies with that predicted by Kepler's laws. While both methods entailed large uncertainties, the dynamical method led to the hypothesis that, on the scale of galaxies and galaxy clusters, a large portion of matter takes the form of dark matter, i.e., is detectable only by its gravitational effect (Ostriker and Mitton 2013 pp 174-197).

An important new approach to measuring the density of matter was suggested in the mid-1950s, namely to search for an expected slowing in the rate of cosmic expansion over time (Longair 2006 p118). From the Friedmann equations, it is easily shown that for models without a cosmological constant, one might expect a slowing in the rate of cosmic expansion due to the presence of matter; indeed a measurement of the de-acceleration parameter $q_0$ could yield a direct estimate of the density of matter. Most importantly, a value of $q_0 = -1$ was predicted for steady-state models, raising the prospect of a clear distinction between steady-state and 'big bang' cosmologies based on measurements of the Hubble constant at different epochs (Hoyle and Sandage 1956). By the end of the decade, astronomical observations were beginning to favour evolving models of the cosmos; these findings were to be bolstered by many new observations of cosmological significance in the 1960s, from observations of the distribution of galaxies over different epochs to the detection of the cosmic microwave background in 1965. Sadly, Einstein did not live to witness these cosmological vindications of the general theory of relativity; however, the year of his death saw the first international conference on general relativity and gravitation, marking the beginning of the renaissance of his great masterwork.

## 5. Conclusions

Although the confrontation of general relativity with experiment was originally framed in terms of three classical astronomical tests proposed by Einstein in 1915, the theory soon found important application in the field of cosmology. Indeed, Einstein's 1917 relativistic model of the cosmos established the basic framework for modern cosmology. By the early 1930s, the first observational evidence of an expansion on cosmic scales had trigged a paradigm shift to dynamic models of the cosmos, almost all of which were couched firmly within the framework of the general theory of relativity. By the end of the 1950s, the attempt by astronomers to determine accurate estimates of key parameters of relativistic models of the cosmos such as the rate of expansion, the curvature of space and the mean density of matter was well under way, setting the stage for today's precision cosmology.

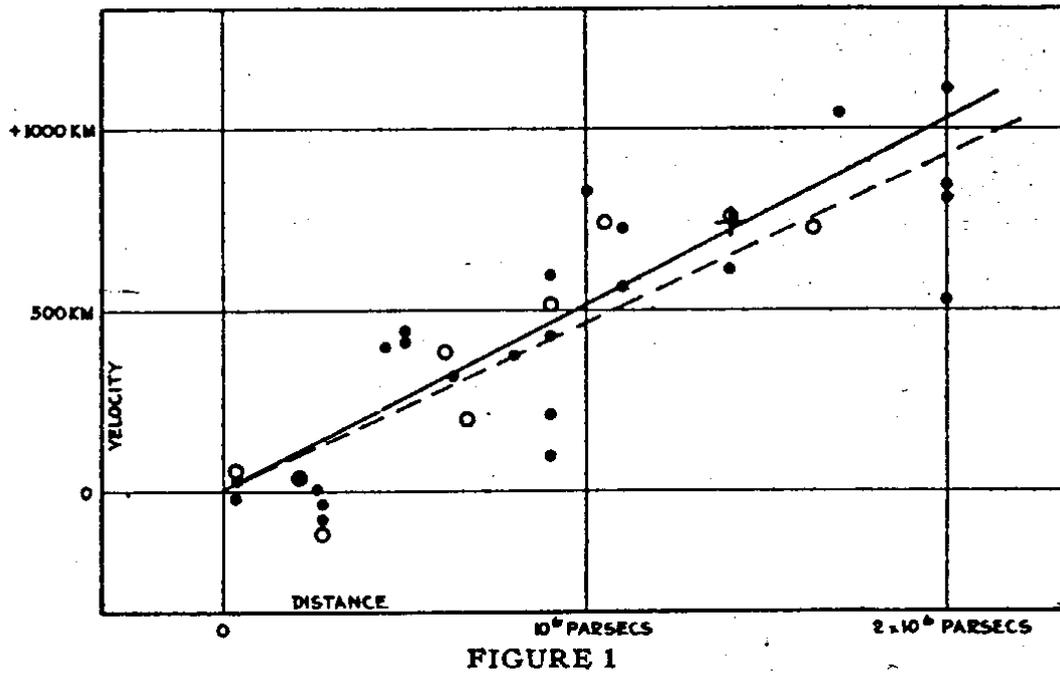

**Figure 1.** Graph of radial velocity vs distance for the spiral nebulae (reproduced from Hubble 1929). Open circles represent data where solar motion was corrected for nebulae in groups; filled circles represent data where solar motion was corrected for individual nebulae.

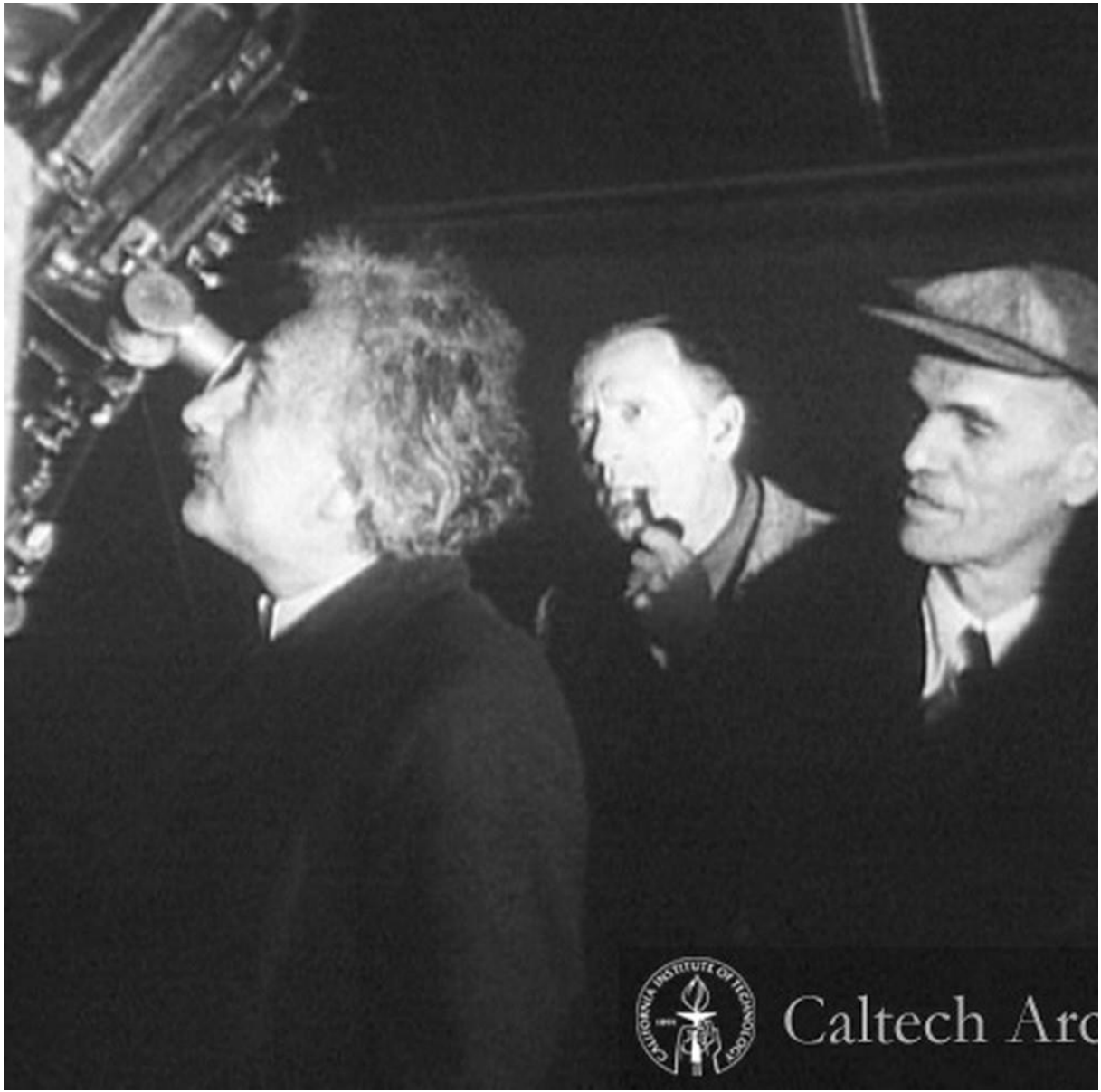

**Figure 2**. Einstein with Edwin Hubble and Walter Adams at the Mount Wilson Observatory in California in 1931.